\documentclass[reprint, twocolumn, twoside,  nofootinbib, hidelinks, colorlinks, linkcolor=blue, citecolor=blue, urlcolor=black,
 amsmath,amssymb,
 aps, pre, %showpacs, showkeys,
floatfix,
]{revtex4-1}

\usepackage[switch,columnwise]{lineno}
\usepackage{fancyhdr}
\usepackage{helvet}
\usepackage{graphicx}% Include figure files
\usepackage{dcolumn}% Align table columns on decimal point
\usepackage{bm}% bold math
\usepackage{hyperref}% add hypertext capabilities
\usepackage{mathrsfs}
\usepackage{amsbsy}
\usepackage{amsmath}
\usepackage{scrextend}
\begin{document}

\preprint{APS/123-QED}

\title{Moving Manifolds in Electromagnetic Fields}% Force line breaks with \\

\author{David V. Svintradze}
 \email{david.svintradze@tsu.ge; dsvintra@yahoo.com}
\affiliation{%
 Department of Physics, Tbilisi State University, Chavchavadze Ave. 03, Tbilisi 0179, Georgia
}%

\date{\today}% It is always \today, today,

\begin{abstract}
We propose dynamic non-linear equations for moving surfaces in electromagnetic field. The field is induced by a material body with a boundary of the surface. Correspondingly the potential energy, set by the field at the boundary, can be written as an addition of four-potential times four-current to a contraction of electromagnetic tensor. Proper application of minimal action principle to the system Lagrangian yields dynamic non-linear equations for moving three dimensional manifolds in electromagnetic fields. The equations, in different conditions simplify to Maxwell equations for massless three surfaces, to Euler equations for dynamic fluid, to magneto-hydrodynamic equations and to Poisson-Boltzmann equation. To illustrate effectiveness of the equations of motion we apply the formalism to analyze dynamics of macro-molecules and membranes. 
\end{abstract}

%\pacs{47.63.-b, 82.70.Uv, 68.15.+e, 68.55.-a.}% PACS, the Physics and Astronomy
%\keywords{Moving Manifolds, Electromagnetic Field, Hydrophobic and Hydrophilic Interactions}%Use showkeys class option if keyword
\maketitle
%\tableofcontents

\section{introduction}

Fluid dynamics is one of the most well understood subjects in classical physics \cite{landau1959} and yet continues to be an actively developing field of research even today. Fluid dynamics can be treated as a motion of an inviscid fluid, as an indivisible medium of particles or as a collective motion of many body system particles. In the first case, when the fluid is inviscid and indivisible, the two conditions allow formulation of the Euler equation for dynamic fluid and the equation of continuity, where the Euler equation is a direct consequence of Newton's second law \cite{landau1959}.  The second case is the most complicated and is difficult to treat. There are two possibilities for dealing with the second case: treat each separate particle as an individual one and propose that each particle satisfies Newton's laws of motion,\footnote{If one proposes to treat particles as classical objects, then the framework fits in Newtonian mechanics. The application of Newton laws and it's stochastic generalizations in simulations is commonly known as molecular dynamics simulations.} or treat each particle as a vertex of a geometric figure and search for equations of motion for such geometries. If smoothed, such geometries for a sufficient number of particles can be modeled as continuously differentiable two manifolds embedded in Euclidean space (classical limit), or continuously differentiable three manifolds embedded in Minkowskian space-time (relativistic limit). Discussion of fluid dynamics in Minkowskian space-time corresponds to the fully relativistic formulation of the problem, while fluid dynamics in Euclidean space corresponds to the non-relativistic limit and is a specific case. 

An example of fluid dynamics modeling as moving surfaces embedded in Euclidian space is moving two dimensional surfaces of fluid films such as soap films. Another, biologically relevant examples are dynamic fluid membranes, vesicles and micelles where large body of notable theoretical results had already been produced \cite{seifert1997, lipowsky1995}.
    
Soap films can be formed by dipping a closed contour wire or by dipping two rings into the soapy solution. Stationary fluid films, or films in mechanical equilibrium with the environment, form a surface with minimal surface area. Usually surfaces such as soap films are modelled as two dimensional manifolds. Fluid films not in mechanical equilibrium may have large displacements and can undergo big deformations \cite{drenckhan2008, brazovskaia1996, kraus1998, seychelles2008, grinfeld2010, manor2015}. The order of magnitude of thickness variations may vary from the nanometer to millimeter scale.

The equations of motion for free liquid films were initially proposed in \cite{grinfeld2009} based on the least action principle of the Lagrangian: 
\begin{equation}
L=\frac{1}{2} \int_S {\rho {(C^2+V^2)}} dS - \sigma \int_S dS \label{1}
\end{equation}
where $\rho$ is the two dimensional mass density of the fluid film, $C$ is interface velocity, $V$ is tangential velocity, $\sigma$ is surface tension, $S$ stands for the surface and free means that interactions with ambient environment are ignored. Numerical solutions of the dynamic nonlinear equations for free thin fluid films display a number of new features consistent with experiments \cite{grinfeld2010}.

As indicated above, fluid dynamics can be described by motion of fluid surfaces, where the motion can happen in Euclidean ambient space, corresponding to the non-relativistic case or in Minkowski ambient space, corresponding to the fully relativistic case. Minkowskian space-time is more general and we will carry out derivations in Minkowski space that can be trivially simplified for non-relativistic cases. Instead of motion of free fluid films, we discuss motion of charged or partially charged material bodies with the boundary of charged or partially charged surfaces\footnote{e.g. bio-membranes, macromolecular surfaces, lipid bilayers, micelles, etc.} in aqueous solution making hydrophobic-hydrophilic interactions. Hydrophobic-hydrophilic interactions are represented as electromagnetic interactions for reasons explained below. Representation of surfaces requires physical modeling and is illustrated in the physical models subsection for biomacromolecular surfaces. To be applicable to biological problems, we take the environment to be aqueous solution, though the medium does not directly enter into the general equations for free moving surfaces, so the equations can be applied to any moving surfaces in electromagnetic field. We propose in this paper the modeling of fluid dynamics as moving surfaces in an electromagnetic field and consequently show that this concept non-trivially generalizes classical fluid dynamics. We pursue fully relativistic calculations because for instance for biological macromolecules, femtosecond observations revealed that surface deformations, induced by dynamics of hydration at the surface or by charge transfer for proteins or DNA, usually happening from angstrom to nanometer scale, may occur as fast as from femtosecond to picosecond \cite{pal2002feftosecond,wan2000feftosecond}. This sets upper limit for the interface velocity as high as $C\sim nm/fs=10^6m/s$ and should be incorporated in fully relativistic framework.\footnote{The formalism should be relativistic not only because that relativistic calculations are more general then classical calculations, or proper electrodynamics description requires relativistic frame work, but because that the molecular surface dynamics can be very fast \cite{pal2002feftosecond,wan2000feftosecond}.}

The theoretical concept of hydrophobicity is already developed \cite{lum1999hydrophobicity, chandler2009dynamics} and is used to simulate shape dependence on hydrophobic interactions \cite{patel2011, rotenberg2011, patel2012, limmer2013}.  Although the basic principles of the hydrophobic effect are qualitatively well understood, only recently have theoretical developments begun to explain quantitatively many features of the phenomenon \cite{chandler2005}.

Hydrophobic and hydrophilic interactions can be described as dispersive interactions between permanent or induced dipoles and ionic interactions throughout the molecules \cite{chandler2005, leikin1993}. Unification of all these interactions in one is the electromagnetic interaction's dependence on the interacting body's geometries \cite{svintradze2010, svintradze2015, svintradze2013, svintradze2009}. To lay a foundation for the description of such geometric dependence, we give exact nonlinear equations governing geometric motion of the surface in an electromagnetic field set up by dipole moments of water molecules and partial charges of various molecules.

In the paper we discuss motion of compact and closed manifolds induced by electromagnetic field, where the field is generated by a continuously distributed charge in the material body. The boundary of the body is a semi-permeable surface (manifold) with a charge (or partial charge) and the charge can flow through the surface. Since, the charge in general is heterogeneously distributed in the body, the charge flow induces time variable electromagnetic field on the surface of the body, forcing the motion of the manifold. Consequently, the problem is to find an equation of motion of moving manifolds in the electromagnetic field. Actuality of the problem may be connected to many physics sub-fields, for instance fluid dynamics, membrane dynamics or molecular surface dynamics. For instance the surface of macromolecules in aqueous solutions is permeable to some ions and water molecules and the charge on the surface is heterogeneously distributed. Flow of some ions and water molecules through the surface and uneven distribution of the charge in the macromolecules induces the surface dynamics. Same happens to biological membranes, vesicles, micelles and etc. Here we deduce general partial differential equations of moving manifolds in electromagnetic field and demonstrate that the equations, in different conditions, simplify to Euler equation for fluid dynamics, Poisson-Boltzmann equation for describing the electric potential distribution on the surface and Maxwell equations for electrodynamics.

The formalism presented in this paper can be easily extended to hypersurfaces of any dimension.  The limitation by three surfaces embedded in four space-time, which is necessary to describe electromagnetism \cite{vanderlinde2004}, is a consequence of specificity of the processes that take place on macromolecular surfaces. The surface of macromolecules in aqueous solutions is permeable to some ions and water molecules and the charge on the surface is heterogeneously distributed. Time frame for dynamics of water molecules on the surface can be of femtosecond range. Therefor the surface can be charged with variable mass and charge densities and is continuously deformable. Mathematically the problem formulates as: find equations of motion in electromagnetic field for a closed, continuously differentiated and smooth two dimensional manifold in Euclidean space (non-relativistic case) or three manifolds in Minkowski space-time (relativistic case). Dynamics of the surfaces under the influence of potential energy arises from four-potential time four-current and contraction of the electromagnetic tensor. Kinetic energy of the manifolds is calculated according to the calculus of moving surfaces \cite{grinfeld2010book}. Potential energy set by the object is modeled by the electromagnetic tensor in the same way as for Maxwell’s equations.  Definition of the Lagrangian \cite{svintradze2015} by subtracting potential energy from the kinetic energy and setting the minimum action principal yields nonlinear equations for moving surfaces in electromagnetic field. 

Since Minkowskian space-time does not follow Riemannian geometry, we need a small adjustment of definitions. For Minkowski space-time, which fits to pseudo-Riemannian geometry, we need definitions of arbitrary base pairs of ambient space, even though the definitions look exactly same as for Riemannian geometry embedded in Euclidean ambient space \cite{grinfeld2010book, weyl1952, svintradze2016}. The summarized relationships about Riemannian geometry embedded in Euclidean space are given in tensor calculus text books \cite{grinfeld2010book, weyl1952}.

\section{theoretical preliminaries}
\subsection{Embedded manifolds in ambient Minkowksi space}

Combination of three ordinary dimensions with the single time dimension forms a four-dimensional manifold and represents Minkowski space-time. In this framework Minkowski four-dimensional space-time is the mathematical model of physical space in which Einstein’s general theory is formulated. Minkowski space is independent of the inertial frame of reference and is a consequence of the postulates of special relativity \cite{weyl1952, landau1971}.

Euclidean space is the flat analog of Riemannian geometry while Minkowski space is considered as the flat analog of curved space-time, which is known in mathematics as pseudo-Riemannian geometry. Considerations of four-dimensional space-time makes embedded moving manifolds three dimensional, where parametric time $t$, describing motion of manifolds, may not have anything to do with proper time $\tau$ used in general relativity.

To briefly describe Minkowskian space-time, let us refer to arbitrary coordinates $X^\alpha$, $\alpha=0,...,3$, where the position vector $\bm{R}$ is expressed in coordinates as $\bm{R}=\bm{R}(X^\alpha)$. Bold letters, throughout the manuscript designate vectors. Latin letters in indexes indicate surface related tensors. Greek letters in indexes show tensors related to the ambient space. All equations are fully tensorial and follow the Einstein summation convention. 

\begin{figure}
\includegraphics{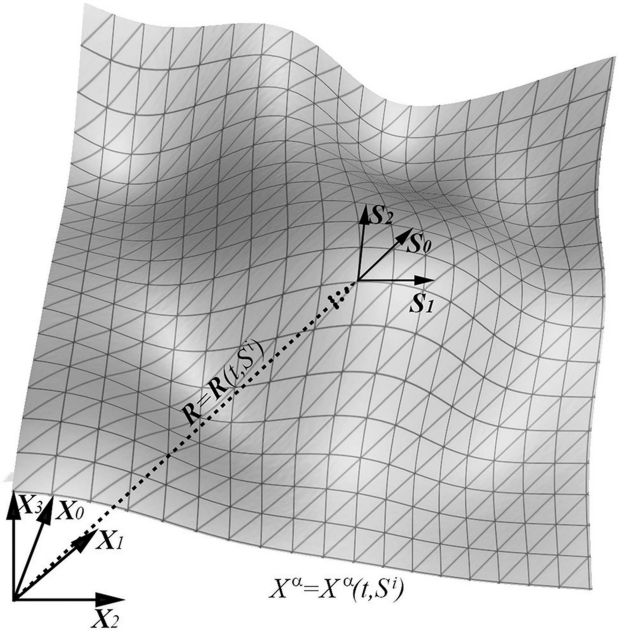}
\caption{\label{fig:1} Two dimensional illustration of a curved three dimensional surface embedded in Minkowski space-time. $X^\alpha$ represents analog of Cartesian coordinates of Minkowski space-time. $\bm{S}_i$ is base vectors defined in tangent space and $X^\alpha=X^\alpha (t,S^i)$ is the general equation of the surface.}
\end{figure}

Suppose that $S^i$ ($i=0,1,2$) are the surface coordinates of the moving manifold $S$ (Fig.~\ref{fig:1}).
Coordinates $S^i$, $X^\alpha$ are arbitrary chosen so that sufficient differentiation is achieved in both space and parametric time. The surface equation in ambient coordinates can be written as ${X^\alpha=X^\alpha(t,S^i)}$ and the position vector can be expressed as
\begin{equation}
\bm{R}=\bm{R}(X^\alpha)=\bm{R}(t,S^i) \label{2}
\end{equation}
Covariant bases for the ambient space are introduced as ${\bm{X}_\alpha=\partial_\alpha\bm{R}}$, where $\partial_\alpha=\partial/{\partial{X^\alpha}}$. The covariant metric tensor is the dot product of covariant bases
\begin{equation}
\eta_{\alpha\beta}=\bm{X}_\alpha\cdot{\bm{X}_\beta} \label{3}
\end{equation}
The contravariant metric tensor is defined as the matrix inverse of the covariant metric tensor, so that ${\eta^{\alpha\nu}\eta_{\nu\beta}=\delta^\alpha_\beta}$, where $\delta^\alpha_\beta$ is the Kronecker delta. From definition (\ref{3}) follows  that $\eta_{00}=\bm{X}_0\cdot\bm{X}_0$ consequently if for Minkowskian space-time, the space like signature is set $(-1,+1,+1,+1)$, then ${\bm{X}_0=(i,0,0,0)}$.\footnote{here the speed of light is set to be unit $c=1$ and $i$ stands for imaginary number.} Therefore, vector components are complex numbers in general. As far as the ambient space is set to be Minkowskian, the covariant bases are linearly independent, so that the square root of the negative metric tensor determinant is unit $\sqrt{-|\eta_{..}|}=1$. Furthermore, the Christoffel symbols given by
\begin{equation}
{\Gamma_{\beta\gamma}^\alpha=\bm{X}^\alpha\cdot\partial_\beta\bm{X}_\gamma} \nonumber
\end{equation}  
vanish and the equality between partial and curvilinear derivatives follows $\partial_\alpha=\nabla_\alpha$. As far as, in Minkowski space-time (later space), $\partial_\alpha$ partial derivative  and $\nabla_\alpha$ curvilinear derivative  are the same, everywhere in calculations we use $\partial$ letter for the ambient space derivative and keep in mind that when referring to Minkowski space the derivative has index in Greek letters and, in that case, it is same as partial derivative. When indexes are mixed Greek and Latin letters the last statement, as it is shown below, does not hold in general.

Now let's discuss tensors on the embedded surface with arbitrary coordinates $S^i$, where $i=0,1,2$. Latin indexes throughout the text are used exclusively for curved surfaces and curvilinear derivative $\nabla_i$ is no longer the same as the partial derivative $\partial_i=\partial/{\partial{S^i}}$. Similar to the bases of ambient space, covariant bases of an embedded manifold are defined as $\bm{S}_i=\partial_i\bm{R}$ and the covariant surface metric tensor is the dot product of the covariant surface bases:
\begin{equation}
g_{ij}=\bm{S}_i\cdot\bm{S}_j \label{4}
\end{equation}
The definition (\ref{4}) dictates that the surface is three dimensional pseudo Riemannian manifold, because ambient space is four dimensional Minkowskian space and the surface in four manifold is three manifold.

analogically to space metric tensor, $g^{ij}$ the contravariant surface metric tensor is the matrix inverse of the covariant one $g_{ij}$. The matrix inverse nature of covariant-contravariant metrics gives possibilities to raise and lower indexes of tensors defined on the manifold. The surface Christoffel symbols are given by
\begin{equation}
 {\Gamma_{jk}^i=\bm{S}^i\cdot\partial_j\bm{S}_k} \nonumber
 \end{equation}  
and along with Christoffel symbols of the ambient space provide all the necessary tools for covariant derivatives to be defined as tensors with mixed space/surface indexes:
\begin{align}
\nabla_i T_{\beta k}^{\alpha j}=&\partial_i T_{\beta k}^{\alpha j}+
\eta_i^\gamma\Gamma_{\gamma\nu}^\alpha T_{\beta k}^{\nu j}-
\eta_i^\gamma\Gamma_{\gamma\beta}^\mu T_{\mu k}^{\alpha j} \nonumber \\ 
&+\Gamma_{im}^jT_{\beta k}^{\alpha m}-\Gamma_{ik}^mT_{\beta m}^{\alpha j} \label{5}
\end{align}
where $\eta_i^\gamma$ is the shift tensor which reciprocally shifts space bases to surface bases, as well as space metric to surface metric. For instance, $\bm{S}_i=\eta_i^\alpha\bm{X}_\alpha$ and
\begin{equation}
{g_{ij}=\bm{S}_i\cdot\bm{S}_j=\eta_i^\alpha\bm{X}_\alpha\eta_j^\beta\bm{X}_\beta=\eta_i^\alpha\eta_j^\beta \eta_{\alpha\beta}} \nonumber
\end{equation}

Metrilinic property $\nabla_ig_{mn}=0$ of the surface metric tensor is direct consequence of (\ref{4},\ref{5}) definitions, therefore ${\bm{S}_m\cdot\nabla_i\bm{S}_n=0}$. The $\bm{S}_m$ and $\nabla_i\bm{S}_n$ vectors are orthogonal, so that  $\nabla_i\bm{S}_n$ must be parallel to $\bm{N}$ surface normal
\begin{equation}
\nabla_i\bm{S}_j=\bm{N} B_{ij} \label{6}
\end{equation}
where $B_{ij}$ is the tensorial coefficient of the (\ref{6}) relationship and is generally referred as the symmetric curvature tensor. The trace of the curvature tensor with upper and lower indexes is the mean curvature and its determinant is the Gaussian curvature. It is well known that a surface with constant Gaussian curvature is a sphere, consequently a sphere can be expressed as:
\begin{equation}
B_i^i=const \label{7}
\end{equation}
When the constant becomes null the surface becomes either a plane or a cylinder. (\ref{7}) is the expression of constant mean curvature (CMC) surfaces in general. Finding the curvature tensor defines the way of finding covariant derivatives of surface base vectors and as so, (\ref{6},\ref{7}) provide the way of finding surface base vectors which indirectly leads to the identification of the surface.

\subsection{Differential geometry for embedded moving manifolds}

After defining metric tensor for ambient space $\eta_{\mu\nu}$ (\ref{3}) and metric tensor for moving surface $g_{ij}$ (\ref{4}) we now proceed with brief review of surface velocity, $t$ explicit (parametric) time derivative of surface tensors and time differentiation theorems for the surface/space integrals. The original definitions of time derivatives for moving surfaces were given in \cite{hadamard1968} and recently extended in \cite{grinfeld2010book}.

\begin{figure}
\includegraphics{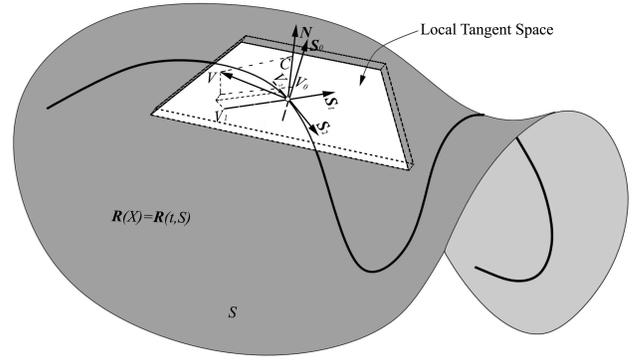}
\caption{\label{fig:velocity} 2D Graphical illustration of the arbitrary chosen three manifold and it's local tangent space. $\bm{S}_0$, $\bm{S}_1$, $\bm{S}_2$ and $\bm{N}$ are local tangent space base vectors and the normal respectively. $\bm{V}$ is arbitrary chosen surface velocity and $C$, $V_i, i=(0,1,2)$ display the projection of the velocity to the $\bm{N},\bm{S}_i$ directions.}
\end{figure}

For the definition of surface velocity we need to define ambient coordinate velocity $V^\alpha$ first and to show that the coordinate velocity is $\alpha$ component of the surface velocity. Indeed, by the velocity definition 
\begin{equation}
V^\alpha=\frac{\partial X^\alpha}{\partial t}
\end{equation}
Taking into account (\ref{2}), the position vector $\bm{R}$ is tracking the material point coordinate $S^i$, therefore by the partial time differentiation of (\ref{2}) and definition of ambient base vectors, we find that $\bm{V}$ surface velocity is
\begin{equation}
\bm{V}=\frac{\partial \bm{R}(t,S^i)}{\partial t}=\frac{\partial \bm{R}}{\partial X^\alpha}\frac{\partial X^\alpha(t,S^i)}{\partial t}=V^\alpha\bm{X}_\alpha \label{9}
\end{equation}
Consequently $V^\alpha$ is ambient component of the surface velocity. According to (\ref{9}), normal component of the surface velocity is the dot product with the surface normal
\begin{equation}
C=\bm{V}\cdot\bm{N}=V_\alpha\bm{X}^\alpha N^\beta\bm{X}_\beta=V_\alpha N^\beta\eta_\beta^\alpha=V_\alpha N^\alpha \label{10}
\end{equation}
The normal component $C$ of the surface velocity is generally referred as an interface velocity and is invariant in contrast with coordinate velocity $V^\alpha$. Its sign depends on a choice of the normal. The projection of the surface velocity on the tangent space (Fig.~\ref{fig:velocity}) \cite{svintradze2016} is tangential velocity and can be expressed as
\begin{equation}
V^i=V^\alpha \eta_\alpha^i \label{11}
\end{equation}
Graphical illustrations of coordinate velocity $V^\alpha$, interface velocity $C$ and tangential velocity $V^i$ are given on Fig.~\ref{fig:velocity}.
There is a clear geometric interpretation of the interface velocity \cite{grinfeld2010book, svintradze2016} and can be expressed as
\begin{equation}
\bm{V}=C\bm{N}+V^i\bm{S}_i \label{12}
\end{equation}
Let the surface at two nearby moments of time $t$ and $t+\Delta t$ be $S_t$, $S_{t+\Delta t}$ correspondingly. Suppose that $A$ is a point on the $S_t$ surface and the corresponding point $B$, belonging to the $S_{t+\Delta t}$, has the same surface coordinate as $A$ (Fig.~\ref{fig:interface velocity}), then ${\bm{AB}\approx\bm{V}\Delta t}$. Let $P$ be the point, where the unit normal $\bm{N} \in S_t$ intersects the surface $S_{t+\Delta t}$, then for small enough $\Delta t$, the angle ${\angle APB\approx\pi/2}$ and $AP\approx\bm{V}\cdot\bm{N}\Delta t$, therefor, $C$ can be defined as 
\begin{equation}
C=\lim_{\Delta t\rightarrow 0}\frac{AP}{\Delta t} \nonumber
\end{equation}
and can be interpreted as the instantaneous velocity of the interface in the normal direction. It is worth of mentioning that the sign of the interface velocity depends on the choice of the normal. Although $C$ is a scalar, it is called interface velocity because the normal direction is implied.

\subsection{Time derivative}

In this section we are briefly explaining concept behind the invariant time derivative for scalar and tensor fields defined on moving manifolds, even though these concepts are already given \cite{grinfeld2010book}. 
\begin{figure}
\includegraphics{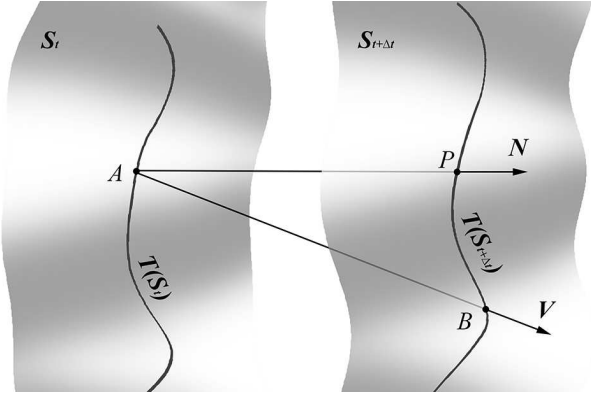}
\caption{\label{fig:interface velocity} Geometric interpretation of the invariant time derivative $\dot{\nabla}$ applied to invariant tensor field $T$. $A$ is an arbitrary chosen point on the $S_t$ surface so that it lays on the $T(S_t)$ curve. $B$ is corresponding point on the $S_{t+\Delta t}$ surface. $P$ is the point where $S_t$ surface normal, applied on the point $A$, intersects the surface $S_{t+\Delta t}$. For infinitely small $\Delta t$ $\bm{AB}\approx\bm{V}\Delta t$ and $AP\approx\bm{V}\cdot\bm{N}\Delta t$. According to geometric construction the tensor field $T$ in the point $B$ can be estimated as $T(B)\approx T(A)+\Delta t\partial T/\partial t$, while $T(B)$ can be estimated by covariant surface derivative $T(B)\approx T(P)+\Delta t V^i\nabla_iT$, where $\nabla_iT$ shows rate of change in $T$ along the directed distance $BP\approx\Delta t V^i$ on the surface $S_{t+\Delta t}$.}
\end{figure}
Suppose that invariant tensor field $T$ is defined on the manifold at all time. To define time invariant derivative of the tensor field it is necessary to capture the rate of change of $T$ in the normal direction. Physical explanation of why the deformations along the normal direction are so important, we give below when discussing integrals. This is similar to how $C$ measures the rate of deformation in the normal direction. Let for a given point $A\in S_t$, find the point $B\in S_{t+\Delta t}$ and $P$ the intersection of $S_{t+\Delta t}$ and the straight line orthogonal to $S_t$  (Fig.~\ref{fig:interface velocity}). Then the geometrically intuitive definition dictates that
\begin{equation}
 \dot{\nabla}T=\lim_{\Delta t\rightarrow 0}\frac{T(P)-T(A)}{\Delta t} \label{13}
 \end{equation} 
because of (\ref{13}) is entirely geometric, it mast be free from choice of a reference frame, therefore it is invariant. On the other hand, from the geometric construction follows that
\begin{equation}
T(B)\approx T(A)+\Delta t\frac{\partial T}{\partial t} \label{14}
\end{equation}
$T(B)$ is related to $T(P)$ because $B$,$P$ are nearby points and are situated on the $S_{t+\Delta t}$ surface $B,P\in S_{t+\Delta t}$, then
\begin{equation}
T(B)\approx T(P)+\Delta tV^i\nabla_iT \label{15}
\end{equation}
since $\nabla_iT$ shows rate of change in the tensor field along the surface and $\Delta tV^i$ indicates the directed distance $BP$. After few lines of algebra, taking into account equations (\ref{14},\ref{15}) in (\ref{13}), we find
\begin{equation}
\dot{\nabla}T=\frac{\partial T}{\partial t}-V^i\nabla_iT \label{16}
\end{equation}
Generalization of (\ref{16}) to any arbitrary tensors with mixed space and surface indexes
is given by the formula
\begin{align}
\dot{\nabla}T_{\beta j}^{\alpha i}=& \frac{\partial T_{\beta j}^{\alpha i}}{\partial t}-V^k\nabla_kT_{\beta j}^{\alpha i}+V^\gamma\Gamma_{\gamma\mu}^\alpha T_{\beta j}^{\mu i}-V^\gamma\Gamma_{\gamma\beta}^\mu T_{\mu j}^{\alpha i} \nonumber \\
&+\dot{\Gamma}_k^iT_{\beta j}^{\alpha k}-\dot{\Gamma}_j^kT_{\beta k}^{\alpha i} \label{17}
\end{align}  
Where Christoffel symbol $\dot{\Gamma}_m^n$ for moving surfaces is ${\dot{\Gamma}_m^n=\nabla_mV^n-CB_m^n}$. The derivative commutes with contraction, satisfies sum, product and chain rules, is metrinilic with respect to the ambient metrics and does not commute with the surface derivative \cite{grinfeld2010book}. Also from (\ref{13}) it is clear that the invariant time derivative applied to time independent scalar vanishes. 

\subsection{Time derivatives of space/surface integrals}

In evaluation of the least action principal of the Lagrangian there is a central role for time differentiation of the surface and space integrals, from which the geometry dependence of the potential energy is rigorously clarified. For any scalar field $T=T(t,S^i)$ defined on a Minkwoskian domain $\Omega$ with boundary $S$ manifold evolving with the interface velocity $C$, the evolution of the space integral and surface integral for closed compact manifolds are given by the formulas
\begin{align}
\frac{d}{dt}\int_\Omega Td\Omega &=\int_\Omega\frac{\partial T}{\partial t}d\Omega+\int_SCTdS \label{18} \\
\frac{d}{dt}\int_S TdS &=\int_S\dot{\nabla}TdS-\int_SCTdS \label{19}
\end{align}
The first term in the integral represents the rate of change of the tensor field, while the second term shows changes in the geometry, therefore properly takes into account the convective and advective terms due to volume motion. We are not going to reproduce proof of these theorems here,\footnote{proofs about the time derivative of integrals can be found in tensor calculus text books, see for instance \cite{grinfeld2010book} and references therein.} but instead we give intuitive explanation of why only interface velocity has the role and tangential velocities do not appear in the integration. If the surface velocity has no interface velocity and has only tangential components, then the tangent velocity translates each points to it's neighboring ones and does not add new area and volume to the surface and space. Therefore, it provokes rotational movement of the material object and can be excluded from the integration. This statement becomes obvious for one dimensional motion. If the material point is moving along some trajectory, then the velocity is tangential to the curve. However, motion of the material point along the curve can be understood as the motion of the curve embedded in the plane. If the curve has no interface velocity then it only slides in the ambient plane without changing the local length.\footnote{same explanation, with more details, is given in \cite{svintradze2016}.}

\subsection{Several useful theorems}

In this section we provide several theorems, which will be directly used to deduce equations of motions. First such theorem is general Gauss theorem about integration, which gives the rule to vise verse  transfer space integral to surface integral. For a domain $\Omega$ in Minkowski space with the boundary $S$, for any sufficiently smooth tensor field $T^\alpha$, the Gauss theorem reads
\begin{equation}
\int_\Omega\nabla_\alpha T^\alpha d\Omega=\int_SN_\alpha T^\alpha dS \label{20}
\end{equation}  
Proof is pretty simple if one uses Voss-Weyl formula to deduce the theorem. For any sufficiently smooth tensor field in Minkowski space, Voss-Weyl formula \cite{grinfeld2010book} reads
\begin{equation}
\nabla_\mu T^\mu=\frac{1}{\sqrt{-|\eta_{..}|}}\partial_\mu(\sqrt{-|\eta_{..}|}T^\mu) \label{21}
 \end{equation} 
Using (\ref{21}) in right part of (\ref{20}) and designation $\eta=-|\eta_{..}|$,  we have
\begin{align}
\int_\Omega\nabla_\alpha T^\alpha d\Omega &=\int_\Omega\frac{1}{\sqrt{\eta}}\partial_\mu(\sqrt{\eta}T^\mu)d\Omega \nonumber \\
&=\iiiint\frac{1}{\sqrt{\eta}}\partial_\mu(\sqrt{\eta}T^\mu)\sqrt{\eta}dX^\alpha \nonumber \\
&=\iiiint\partial_\mu(\sqrt{\eta}T^\mu)dX^\alpha \nonumber
\end{align}
where $dX^\alpha=dX^0dX^1dX^2dX^3$. This term is subject to Gauss’s theorem in the arithmetic space. Since, the arithmetic space and Minkowski space, which is a pseudo-Euclidean, can be corresponded to the Cartesian coordinates, the Minkowski space can be identified as arithmetic one and the Gauss theorem for the arithmetic space can be used. Thus, using unity of Minkowski space metric tensor determinant one may prove that\footnote{details about the proof, for Euclidean space, can be found in tensor calculus text book \cite{grinfeld2010book} and proof for Minkowski space is identical to Euclidean one.}
\begin{align}
\iiiint\partial_\mu(\sqrt{\eta}T^\mu)dX^\alpha&=\iiint N_\alpha T^\alpha\sqrt{\eta}\sqrt{g}dS^i \nonumber \\
&=\int_SN_\alpha T^\alpha dS \nonumber
\end{align}
where $g=|g_{..}|$. This proves that generalized Gauss's theorem holds for pseudo-Riemannian manifolds embedded in Minkowski space.

Next step is to provide short proofs for Weingarten's and Thomas formulas by using the relation between the surface derivative and the interface velocity.

Weingarten's formula expresses surface covariant derivative of the surface normal in the product of the shift and mixed curvature tensors. Proof follows from the definition $N_\alpha N^\alpha =1$, from where we find $N_\alpha\nabla_i N^\alpha=0$. On the other hand 
\begin{align}
0=\bm{N}\cdot\bm{S}_i=N^\alpha\bm{X}_\alpha\cdot\eta^\beta_i\bm{X}_\beta &=N^\alpha\eta_i^\beta\eta_{\alpha\beta} \nonumber \\
&=N^\alpha\eta_{\alpha i} \label{22}
     \end{align}     
If we apply covariant derivative to (\ref{22}) and take into account that from (\ref{6}) follows $B_{ji}=N^\alpha\nabla_j\eta_{i\alpha}$  then by the product rule we find
\begin{align}
0=\nabla_jN^\alpha\eta_{\alpha i}+N^\alpha\nabla_j\eta_{\alpha i}&=\nabla_jN^\alpha\eta_{\alpha i}+B_{ji} \nonumber \\
\nabla_jN^\alpha\eta_{\alpha i}&=-B_{ji} \label{23}
\end{align}
Let's contract both sides of (\ref{23}) with $\eta^{i\beta}$ and take into account commonly used relationship in tensor calculus ${N^\alpha N_\beta+\eta_i^\alpha\eta_\beta^i=\delta_\beta^\alpha}$, then we find 
\begin{align}
-\eta^\beta_kB_j^k &=-\eta^{i\beta}B^k_jg_{ki}=-\eta^{i\beta}B_{ji}=\nabla_jN^\alpha\eta_{\alpha i}\eta^{i\beta} \nonumber \\
&=\nabla_jN^\alpha\eta_\alpha^kg_{ik}\eta^\beta_mg^{mi}=\nabla_jN^\alpha\eta_\alpha^k\eta^\beta_m\delta_k^m \nonumber \\
&=\nabla_jN^\alpha\eta_\alpha^k\eta^\beta_k=\nabla_jN^\alpha(\delta_\alpha^\beta-N^\beta N_\alpha) \nonumber \\
&=\nabla_jN^\beta-N^\beta N_\alpha\nabla_jN^\alpha \nonumber \\
\nabla_jN^\beta&=-\eta^\beta_kB_j^k \label{24}
\end{align}
Since the second term of the last equality vanishes, we get (\ref{24}) also known as Weingarten's formula.

Now we turn to the Thomas formula allowing to calculate invariant time derivative of the surface normal. Indeed, using invariant time derivative formula for surface base vector \cite{grinfeld2010book}
\begin{equation}
\dot{\nabla}\bm{S}_i=\bm{N}\nabla_iC \label{25}
\end{equation} 
and doting both sides of (\ref{25}) with $\bm{N}$, ${\bm{N}\cdot\dot{\nabla}\bm{S}_i=\bm{N}\cdot\bm{N}\nabla_iC}$ and using product rule, taking into account that $\bm{N}\cdot\bm{S}_i=0$, we find $\nabla_iC=-\dot{\nabla}\bm{N}\bm{S}_i$, therefor
\begin{equation}
\dot{\nabla}\bm{N}=-\bm{S}^i\nabla_iC \label{26}
\end{equation}
the equation (\ref{26}) is generally referred as Thomas formula.

\section{Equations of Motions and Physical Models}

\subsection{Equations of motions}
Since we have all mathematical preliminaries in hand we can proceed with derivation of master equations of motions. To deduce the equations we apply the calculus of moving surfaces to the motion of manifolds in an electromagnetic field. On this step we only discuss free motion of single surface, where in 'single' surface we mean boundary of the single material body and free means contact with environment is ignored.\footnote{the environment is set to be vacuum.} The interaction with environments can be incorporated into the equations later on.\footnote{in the case of taking into account interaction with environment we no longer have single surface, instead there are double surfaces where first one is the boundary of the material body and another one is the surface of the environment at the boundary/environment interface. Having two surfaces rises the terms related to surface-surface interactions and may enter into final equations as viscoelastic effect incorporated in coefficient of viscosity.} The surface is treated as continuum media of material particles (points), where charge and mass distribution is heterogeneous in general. The boundary of the body is the surface with a surface mass density $\rho$ and a surface charge density $q$. The surface can be semipermeable to some material points, meaning the charge can flow through the surface. Interaction between material points are exclusively electromagnetic, as far as the mass of each material particles are set to be infinitely small comparably to unit charges. As far as the ambient space is set to be Minkowskian, the body is four dimensional and has the surface boundary of three dimensional manifold. Electromagnetic interaction between the material particles and heterogeneous distribution of charges throughout the object induces motion of the surface and the potential energy of the interaction can be modeled as 
\begin{equation}
\mathcal{U}=\int_\Omega(\frac{1}{4\mu_0}F_{\alpha\beta}F^{\alpha\beta}+A_\alpha J^\alpha)d\Omega \label{27}
\end{equation}
where the electromagnetic tensor $F_{\alpha\beta}$ is the combination of the electric and magnetic fields in a covariant antisymmetric tensor \cite{vanderlinde2004, landau1971}. The electromagnetic covariant four-potential is a covariant four vector $\bm{A}_\cdot=(-\varphi/c,\bm{a})$ composed by the $\varphi$ electric potential and the $\bm{a}$ magnetic potential. Contravariant four current $\bm{J}^\cdot=(cQ,\bm{j})$ is the contravariant four vector combining $\bm{j}$ electric current density and $Q$ the charge density, $c$ is a speed of light and $\mu_0$ is the magnetic permeability of the vacuum. Minkowski space metric tensor signature is set to be space-like $(-+++)$ throughout the paper. This formulation is a fully relativistic though it can be easily simplified for non-relativistic cases. Raising and lowering the indexes is performed by the Minkowski metric $\eta_{\alpha\beta}$. The relation between the four potentials and the electromagnetic tensor is given by
\begin{equation}
F_{\alpha\beta}=\partial_\alpha A_\beta-\partial_\beta A_\alpha \label{28}
\end{equation}
As far as the boundary of the material body is moving three manifold, the surface kinetic energy with variable surface mass density $\rho$ and the surface velocity $\bm{V}$ is
\begin{equation}
\mathcal{T}=\int_S\frac{\rho\bm{V}^2}{2}dS \label{29}
\end{equation}
Subtraction of the potential energy (\ref{27}) from the kinetic energy (\ref{29}) leads to the system Lagrangian
\begin{equation}
\mathcal{L}=\int_S\frac{\rho\bm{V}^2}{2}dS-\int_\Omega(\frac{1}{4\mu_0}F_{\alpha\beta}F^{\alpha\beta}+A_\alpha J^\alpha)d\Omega \label{30}
\end{equation}
where $S$ is the boundary of $\Omega$. Hamilton's least action principle \cite{johns2005} for the given Lagrangian (\ref{30}) reads
\begin{equation}
\frac{\delta\mathcal{L}}{\delta t}=\frac{\delta\mathcal{T}}{\delta t}-\frac{\delta\mathcal{U}}{\delta t}=0 \label{31}
\end{equation}
For proper evaluation of the (\ref{31}) Lagrangian we start from the simplest term first, it is variation of the potential energy. Since (\ref{27}) is the space integral by theorem (\ref{18}) we have 
\begin{align}
\frac{\delta\mathcal{U}}{\delta t}=&\int_\Omega\frac{\partial }{\partial t}(\frac{1}{4\mu_0}F_{\alpha\beta}F^{\alpha\beta}+A_\alpha J^\alpha)d\Omega \nonumber \\
&+\int_SC(\frac{1}{4\mu_0}F_{\alpha\beta}F^{\alpha\beta}+A_\alpha J^\alpha) dS \label{32}
\end{align}
According to (\ref{32}) determination of variation of potential energy is to calculate time differential of the space integrand. Following to standard algebraic manipulations for classical electrodynamics, we find 
\begin{align}
\int_\Omega\frac{\partial u}{\partial t} d\Omega =&\int_\Omega(\frac{\partial u}{\partial A_\alpha}\frac{\partial A_\alpha}{\partial t}+\frac{\partial u}{\partial (\partial_\beta A_\alpha)}\frac{\partial(\partial_\beta A_\alpha)}{\partial t})d\Omega \nonumber \\
=&\int_\Omega(\frac{\partial u}{\partial A_\alpha}\frac{\partial A_\alpha}{\partial t}+\partial_\beta(\frac{\partial u}{\partial(\partial_\beta A_\alpha)}\frac{\partial A_\alpha}{\partial t}) \nonumber \\
&-\partial_\beta\frac{\partial u}{\partial(\partial_\beta A_\alpha)}\frac{\partial A_\alpha}{\partial t})d\Omega \nonumber \\
=&\int_\Omega(\frac{\partial u}{\partial A_\alpha}\frac{\partial A_\alpha}{\partial t}-\partial_\beta\frac{\partial u}{\partial(\partial_\beta A_\alpha)}\frac{\partial A_\alpha}{\partial t})d\Omega \nonumber \\
&+{\frac{\partial u}{\partial(\partial_\beta A_\alpha)}\frac{\partial A_\alpha}{\partial t}}|_{\frac{\partial A_\alpha}{\partial t}=0} \nonumber \\
=&\int_\Omega(\frac{\partial u}{\partial A_\alpha}-\partial_\beta\frac{\partial u}{\partial(\partial_\beta A_\alpha)})\frac{\partial A_\alpha}{\partial t}d\Omega \label{33}
\end{align}
where we used $u=(1/4\mu_0)F_{\mu\nu}F^{\mu\nu}+A_\mu J^\mu$ designation and the fact that $u$ is a function of $A_\alpha$ and $\partial_\beta A_\alpha$ and at the boundary condition $\partial A_\alpha/\partial t=0$ the last term vanishes. It is easy to show that,
\begin{equation}
\partial u/\partial A_\alpha=J^\alpha \label{34}
\end{equation}
To calculate the last integrand (\ref{33}), we take into account the definition (\ref{28}) and note that covariant electromagnetic tensor can be obtained by lowering indexes in contravariant tensor $F^{\alpha\beta}=\eta^{\gamma\alpha}\eta^{\kappa\beta}F_{\gamma\kappa}$ and the electromagnetic tensor is antisymmetric $F^{\alpha\beta}=-F^{\beta\alpha}$, so that
\begin{align}
\frac{\partial u}{\partial(\partial_\beta A_\alpha)}=&\frac{1}{4\mu_0}\frac{\partial}{\partial(\partial_\beta A_\alpha)}(F_{\mu\nu}\eta^{\gamma\mu}\eta^{\lambda\nu}F_{\gamma\lambda}) \nonumber \\
=&\frac{1}{4\mu_0}\eta^{\gamma\mu}\eta^{\lambda\nu}\frac{\partial (F_{\mu\nu}F_{\gamma\lambda})}{\partial(\partial_\beta A_\alpha)} \nonumber \\
=&\frac{1}{4\mu_0}\eta^{\gamma\mu}\eta^{\lambda\nu}(F_{\mu\nu}(\delta_\gamma^\beta\delta_\lambda^\alpha-\delta_\lambda^\beta\delta_\gamma^\alpha) \nonumber \\
&+F_{\gamma\lambda}(\delta_\mu^\beta\delta_\nu^\alpha-\delta_\nu^\beta\delta_\mu^\alpha)) \nonumber \\
=&\frac{1}{4\mu_0}(F^{\beta\alpha}-F^{\alpha\beta}+F^{\beta\alpha}-F^{\alpha\beta})=\frac{F^{\beta\alpha}}{\mu_0} \label{35}
\end{align}
Taking into account (\ref{33}-\ref{35}) in (\ref{32}) we find the variation of the potential energy
\begin{align}
\frac{\delta\mathcal{U}}{\delta t}=&\int_\Omega(J^\alpha-\frac{1}{\mu_0}\partial_\beta F^{\beta\alpha})\frac{\partial A_\alpha}{\partial t} d\Omega \nonumber \\
&+\int_SC(\frac{1}{4\mu_0}F_{\alpha\beta}F^{\alpha\beta}+A_\alpha J^\alpha) dS \label{36}
\end{align}

Now we turn to the calculation of the kinetic energy variation. To deduce the variation for the kinetic energy let's define generalization of conservation of mass low first. The variation of the surface mass density must be so that $dm/dt=0$, where
\begin{equation}
m=\int_S\rho dS  \label{37}
\end{equation} 
is the surface mass with $\rho$ surface mass density. Since, we discuss compact closed manifolds the boundary conditions $v=n_iV^i=0$ dictate, that a pass integral along any curve across the surface must vanish. This statement formally, taking into consideration (\ref{37}), can be rewritten as
\begin{align}
0=& \int_\gamma v\rho d\gamma=\int_\gamma n_iV^i\rho d\gamma=\int_S\nabla_i(\rho V^i)dS \nonumber \\
=&\int_S(\nabla_i(\rho V^i)-\rho CB_i^i+\rho CB_i^i)dS \nonumber \\
=&\int_S(\nabla_i(\rho V^i)-\rho CB_i^i)dS+\int_S \dot{\nabla}\rho dS-\frac{d}{dt}\int_S\rho dS \nonumber \\
=&\int_S(\nabla_i(\rho V^i)-\rho CB_i^i+ \dot{\nabla}\rho)dS \label{38}
\end{align}
where $n_i$ is a normal of the curve that lays in the tangent space, $v$ is the velocity of the $\gamma$ curve. Since last integral from (\ref{38}) mast be identical to zero for any integrand, one immediately finds generalization of conservation of mass low
\begin{equation}
\dot{\nabla}\rho+\nabla_i(\rho V^i)=\rho CB_i^i \label{39}
\end{equation}
Incidently, an equation for the surface charge conservetion can be analogically deduced and it has excatly the same form. The equation (\ref{39}) was also reported in \cite{grinfeld2009}. To calculate the variation of the kinetic energy we use (\ref{19},\ref{29},\ref{39}) and after few lines of algebra, we find
\begin{align}
\frac{\delta \mathcal{T}}{\delta t}=&\int_S(\dot{\nabla}\frac{\rho\bm{V}^2}{2}-CB_i^i\frac{\rho\bm{V}^2}{2})dS \nonumber \\
=&\int_S(\dot{\nabla}\rho\frac{V^2}{2}+\rho\dot{\nabla}\frac{V^2}{2}-CB_i^i\frac{\rho V^2}{2})dS \nonumber \\
=&\int_S((\rho CB_i^i-\nabla_i(\rho V^i))\frac{V^2}{2}+\rho\dot{\nabla}\frac{V^2}{2}-CB_i^i\frac{\rho V^2}{2})dS \nonumber \\
=&\int_S(-\nabla_i(\rho V^i\frac{V^2}{2})+\rho V^i\nabla_i\frac{V^2}{2}+\rho\dot{\nabla}\frac{V^2}{2})dS \nonumber \\
=&\int_S\rho\bm{V}(V^i\nabla_i\bm{V}+\dot{\nabla}\bm{V})dS \label{40}
\end{align}
Here we used that at the end of variations the surface reaches the stationary point and, therefore, by Gauss theorem integral for $\nabla_i(\rho V^iV^2/2)$ converted to line integral vanishes (as we used it already in (\ref{38})). To deduce the final form of equations of motion we decompose dot product in the integral (\ref{40}) by normal and tangential components. After few lines of algebraic manipulations, we find
\begin{align}
&\dot{\nabla}\bm{V}+V^i\nabla_i\bm{V}=\dot{\nabla}\bm{V}+V^i\nabla_i\bm{V}+CV^iB_i^j\bm{S}_j-CV^iB_i^j\bm{S}_j \nonumber \\ 
&=\dot{\nabla}\bm{V}+V^i\nabla_i\bm{V}+CV^iB_i^j\eta_j^\alpha\bm{X}_\alpha-CV^iB_i^j\bm{S}_j \label{41} 
\end{align}
Using Weingarten’s formula (\ref{24}), metrinilic property of the Minkowksi space base vectors $\nabla_i\bm{X}_\alpha=0$ and the definition of the surface normal ${\bm{N}=N^\alpha\bm{X}_\alpha}$, the last equation of (\ref{41}) transforms
\begin{align}
&\dot{\nabla}\bm{V}+V^i\nabla_i\bm{V}-CV^i\bm{X}_\alpha\nabla_iN^\alpha-CV^iB_i^j\bm{S}_j \nonumber \\
&=\dot{\nabla}\bm{V}+V^i\nabla_i\bm{V}-CV^i\nabla_i\bm{N}-CV^iB_i^j\bm{S}_j \label{42}
\end{align}
Taking into account (\ref{12}) and its covariant and invariant time derivatives in (\ref{42}), we find
\begin{align}
&\dot{\nabla}\bm{V}+V^i\nabla_i(C\bm{N})+V^i\nabla_i(V^j\bm{S}_j) \nonumber \\
&-CV^i\nabla_i\bm{N}-CV^iB_i^j\bm{S}_j \nonumber \\
&=\dot{\nabla}\bm{V}+V^i\bm{N}\nabla_iC+V^i\nabla_i(V^j\bm{S}_j)-CV^iB_i^j\bm{S}_j \nonumber \\
&=\dot{\nabla}(C\bm{N})+\dot{\nabla}(V^j\bm{S}_j)+V^i\bm{N}\nabla_iC \nonumber \\
&+V^i\nabla_i(V^j\bm{S}_j)-CV^iB_i^j\bm{S}_j \label{43}
\end{align}
Continuing algebraic manipulations using the formula for the surface derivative of the interface velocity (\ref{25}), Thomas formula (\ref{26}) and the definition of the curvature tensor (\ref{6}) in (\ref{43}), yield
\begin{align}
&\dot{\nabla}(C\bm{N})+C\nabla^jC\bm{S}_j+2V^i\bm{N}\nabla_iC+V^iV^jB_{ij}\bm{N} \nonumber \\
&+\dot{\nabla}(V^j\bm{S}_j)-V^i\bm{N}\nabla_iC+V^i\nabla_i(V^j\bm{S}_j)-V^iV^jB_{ij}\bm{N} \nonumber \\
&-C\nabla^jC\bm{S}_j-CV^iB_i^j\bm{S}_j \nonumber \\
&=\dot{\nabla}(C\bm{N})-C\dot{\nabla}\bm{N}+2V^i\bm{N}\nabla_iC+V^iV^jB_{ij}\bm{N} \nonumber \\
&+\dot{\nabla}(V^j\bm{S}_j)-V^j\dot{\nabla}\bm{S}_j+V^i\nabla_i(V^j\bm{S}_j)-V^iV^j\nabla_i\bm{S}_j \nonumber \\
&-C\nabla^jC\bm{S}_j-CV^iB_i^j\bm{S}_j \nonumber \\
&=(\dot{\nabla}C+2V^i\nabla_iC+V^iV^jB_{ij})\bm{N} \nonumber \\
&+(\dot{\nabla}V^j+V^i\nabla_iV^j-C\nabla^jC-CV^iB_i^j)\bm{S}_j \label{44}
\end{align}
Doting (\ref{44}) on $\bm{V}$ and combining it with (\ref{40}) the last derivation reveals variation of the kinetic energy
\begin{align}
\frac{\delta \mathcal{T}}{\delta t}=&\int_S\rho C(\dot{\nabla}C+2V^i\nabla_iC+V^iV^jB_{ij})dS \nonumber \\
&+\int_S\rho V_i(\dot{\nabla}V^i+V^j\nabla_jV^i-C\nabla^iC-CV^jB_j^i)dS \label{45}
\end{align}
where first part is normal component and second part is tangent component of the dot product. Combination of (\ref{36},\ref{45}) with (\ref{31}) reveals
\begin{align}
&\int_S\rho C(\dot{\nabla}C+2V^i\nabla_iC+V^iV^jB_{ij})dS \nonumber \\
&+\int_S\rho V_i(\dot{\nabla}V^i+V^j\nabla_jV^i-C\nabla^iC-CV^jB_j^i)dS \nonumber \\
&=\int_\Omega(J^\alpha-\frac{1}{\mu_0}\partial_\beta F^{\beta\alpha})\frac{\partial A_\alpha}{\partial t} d\Omega \nonumber \\
&+\int_SC(\frac{1}{4\mu_0}F_{\alpha\beta}F^{\alpha\beta}+A_\alpha J^\alpha) dS \label{46}
\end{align} 
To find the final form of equations of motion let's brake dot product of space integrand from (\ref{46}) by normal and tangential components. Let's propose that the vector $\pmb{\mathscr{F}}$  with contravariant $\alpha$ component is
\begin{equation}
\pmb{\mathscr{F}}=(\mathcal{F}^\alpha)=(J^\alpha-\frac{1}{\mu_0}\partial_\beta F^{\beta\alpha})=\mathcal{F}\bm{N}+\mathcal{F}^i\bm{S}_i \label{47}
\end{equation}
where $\mathcal{F}$ and $\mathcal{F}^i$ are normal and tangential components of $\pmb{\mathscr{F}}$ correspondingly. Analogically, for $\partial\bm{A}/\partial t$ four vector partial time derivative, we have
\begin{equation}
\frac{\partial\bm{A}}{\partial t}=\mathcal{A}\bm{N}+\mathcal{A}^i\bm{S}_i \label{48}
\end{equation} 
where $\mathcal{A},\mathcal{A}^i$  are the normal and tangential components of the partial time derivative of the four vector potential. Using the definitions (\ref{47},\ref{48}) the dot product of the two vectors is
\begin{equation}
\pmb{\mathscr{F}}\cdot\frac{\partial\bm{A}}{\partial t}=\mathcal{F}\mathcal{A}+\mathcal{F}^i\mathcal{A}_i \label{49}
\end{equation}
Since the equation (\ref{46}) must hold for every $\bm{V}, \pmb{\mathscr{F}}, \partial\bm{A}/\partial t$ vectors in general, the normal and tangential components of the dot product must be equal so that, taking into account (\ref{47}-\ref{49}) in (\ref{46}), we find
\begin{align}
&\int_S\rho C(\dot{\nabla}C+2V^i\nabla_iC+V^iV^jB_{ij})dS \nonumber \\
&=\int_\Omega\mathcal{F}\mathcal{A}d\Omega +\int_SC(\frac{1}{4\mu_0}F_{\alpha\beta}F^{\alpha\beta}+A_\alpha J^\alpha) dS \label{50} \\
&\int_S\rho V_i(\dot{\nabla}V^i+V^j\nabla_jV^i-C\nabla^iC-CV^jB_j^i)dS \nonumber \\
&=\int_\Omega\mathcal{F}^i\mathcal{A}_id\Omega \label{51}
\end{align}
After applying the Gauss theorem to the surface integrals in (\ref{50}), the surface integrals are converted to space integral so that one gets
\begin{align}
&\int_\Omega\partial_\mu(\rho V^\mu (\dot{\nabla}C+2V^i\nabla_iC+V^iV^jB_{ij}))d\Omega \nonumber \\
&-\int_\Omega\partial_\mu (V^\mu(\frac{1}{4\mu_0}F_{\alpha\beta}F^{\alpha\beta}+A_\alpha J^\alpha))d\Omega=\int_\Omega\mathcal{F}\mathcal{A}d\Omega \label{52}
\end{align}
To summarize (\ref{39},\ref{50}-\ref{52}) equations of moving manifolds in electromagnetic field read
\begin{widetext}
\begin{align}
\dot{\nabla}\rho+\nabla_i(\rho V^i)&=\rho CB_i^i \nonumber \\
\partial_\mu(V^\mu (\rho (\dot{\nabla}C+2V^i\nabla_iC+V^iV^jB_{ij}) 
-\frac{1}{4\mu_0}F_{\alpha\beta}F^{\alpha\beta}-A_\alpha J^\alpha))&=\mathcal{F}\mathcal{A}  \label{53} \\
\int_S\rho V_i(\dot{\nabla}V^i+V^j\nabla_jV^i-C\nabla^iC-CV^jB_j^i)dS
&=\int_\Omega\mathcal{F}^i\mathcal{A}_id\Omega \nonumber
\end{align}
\end{widetext}
Equations (\ref{53}) accomplish the search for master equations of motions. 

Perhaps the cases that deserves some attention are homogeneous symmetrical surface, in that case only nonzero "force" allowed to be is $\mathcal{F}\mathcal{A}\neq 0$ and $\mathcal{F}^i\mathcal{A}_i=0$, this leads to significant simplification of the third equation from (\ref{53}) and second equation can be analytically solved for homogeneous, equilibrium surfaces as we have done it for micelles \cite{svintradze2016}. When $\mathcal{F}\mathcal{A}\neq 0$ and $\mathcal{F}^i\mathcal{A}_i\neq 0$ then motion of the surface induces swimming of the body. The case $\mathcal{F}\mathcal{A}=0$ and $\mathcal{F}^i\mathcal{A}_i\neq 0$, as it is shown below, simplifies to Euler equation for dynamic fluid for free motion and to Navier-Stokes equation or to  magneto-hydrodynamic (MHD) equations if one takes into account interactions with environment. 

Equations (\ref{53}) are correct for freely moving manifolds of the body in the vacuum. Generalization can be trivially achieved if instead of electromagnetic tensor $F_{\alpha\beta}$ one proposes electromagnetic stress energy tensor $T^{\alpha\beta}$, which is related to electromagnetic tensor by relationship
\begin{align}
T^{\alpha\beta}=\frac{1}{\mu_0}(\eta_{\gamma\nu}F^{\alpha\gamma}F^{\nu\beta}+\frac{1}{4}\eta^{\alpha\beta}F_{\gamma\nu}F^{\gamma\nu}) \label{54}
\end{align}  
For objects in matter electromagnetic tensor $F^{\mu\nu}$ in (\ref{47},\ref{53}) is replaced by electric displacement tensor $D^{\mu\nu}$ and by magnetization-polarization tensor $M^{\mu\nu}$ so that
\begin{equation}
\frac{F^{\mu\nu}}{\mu_0}=D^{\mu\nu}+M^{\mu\nu} \label{55}
\end{equation} 
The charge density $Q$ and four current $\bm{J}$ become sum of bound and free charges and of bound and free four currents respectively. Electric displacement tensor, magnetization tensor, free and bound charges/currents can be modeled differently, for different problems,  therefore the general equations (\ref{53}) can be modified as needed.

\subsection{Physical models}

\begin{figure}
\includegraphics{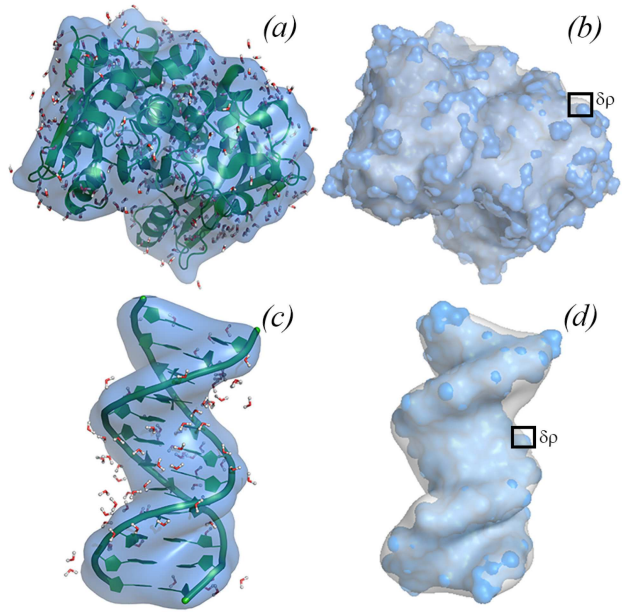}
\caption{\label{fig:protein_DNA} (Color online) Top $(a,b)$: Gaussian map of the protein and Bottom $(c,d)$: of the DNA dodecamer. $(a,c)$ highlights smoothed Gaussian mapping of the macromolecules, at 5 \text{\AA} resolution, indicating crystallographic distribution of the water molecules on the surface. $(b,d)$ shows the mapping of the macromolecules at two different resolutions, capturing surface thickness variation incorporated in $\delta\rho$. Blue surface is mapping at 2 \text{\AA} resolution and indicates distribution of the atoms on the surface, while light gray is smoothed Gaussian map indicating how surface thickness may vary if free diffusion of the solvent molecules on the surface is taken into account.}
\end{figure}

To link above formulated problem with real physical surfaces it is necessary some modeling and for the beginning let's illustrate macromolecules\footnote{or surface made from groups of molecules, for instance lipids} as a two dimensional fluid manifolds with the thickness of variable mass density Fig.~\ref{fig:protein_DNA}. 
Even though molecular surfaces are three manifold in Minkowski space, in some cases\footnote{especially for relatively slowly moving surfaces, for instance: cellular surface, which is bio-membrane; vesicles; micelles etc.} it can be modeled as moving two manifolds in Euclidean space. The surface is considered to be semipermeable against partial charges and water molecules. The permeability defines the surface mass density as a variable and the volume charge also becomes variable. The variability of charge and mass densities is properly taken into account in the equations of motion (\ref{53}).

Let's model a bio-macromolecular surface as a Gaussian map contoured at 2 \text{\AA} to 8 \text{\AA} resolution. Fig.~\ref{fig:protein_DNA} shows Gaussian maps for the protein (Fig.~\ref{fig:protein_DNA}$(a,b)$) and for the DNA (Fig.~\ref{fig:protein_DNA}$(c,d)$). $\Omega$ is the space inside the macromolecules and the boundary of the space is the surface $S$. $\bm{S}_i$ base vectors are defined in the tangent space of the Gaussian map. $S_{ij}$ is the metric tensor of the map. These are illustrations of surfaces as two-manifolds embedded in Euclidian space and are only true for non-relativistic representations, therefore do not show the shape of three-manifolds in Minkowski space-time. Fig.~\ref{fig:protein_DNA} $(a)$ and $(c)$ show a Gaussian map of the polypeptide main chain of a protein and of a polynucleotide double helical DNA dodecamer respectively.\footnote{For a model protein was taken one of a peroxide sensitive gene regulator with Protein Data Bank (PDB) ID 3HO7 \cite{svintradze2013structure} and for model DNA was taken polynucleotide double helical dodecamer generally known as library DNA with PDB ID 1BNA \cite{drew1981}.} Fig.~\ref{fig:protein_DNA} $(b)$ and $(d)$ show thickness variations, captured by surface mass density, of the modeled surfaces for the protein and DNA. Light gray is the Gaussian map at 5 \text{\AA} resolution while the blue surface indicates a more detailed surface contoured at 2 \text{\AA} resolution. Thickness variation can be induced by diffusion of solvent molecules at solvent accessible sites; e.g. sites marked by water molecules obtained from crystal structures as illustrated in the Fig.~\ref{fig:protein_DNA} $(a)$ and $(c)$ (red and white sticks), or by thermal fluctuation of amino acids sidechains. In all these cases, the surface thickness variation, captured by $\rho$ surface mass density, is in the range of angstrom to nanometer. This range is higher for micelles, cell membranes, fluid films etc. If the system is in aqueous solution then the surface motion is determined by so called hydrophobic-hydrophilic interactions.
 
As we already stated in introduction, hydrophobic and hydrophilic interaction incorporate dispersive interactions, throughout the molecules, mainly related to electrostatics and electrodynamics (Van der Waals forces), induced by permanent (water molecules) or induced dipoles (dipole-dipole interactions) and possible quadrupole-quadrupole interactions (for instance stacking or London forces) plus ionic interactions (Coulomb forces) \cite{leikin1993}. The hydrophobic effect can be considered as synonymous with dispersive interactivity with water molecules and the hydrophilic one as synonymous with polar interactivity with water molecules \cite{chandler2009dynamics, chandler2005, leikin1993}. All these interactions have one common feature and can be unified as electromagnetic interaction's dependence on interacting bodies' geometries, where by geometries we mean shape of the objects' surfaces. To model potential energy we note that on the scale of hydrophobic-hydrophilic interactions, which usually occurs at nanometer distances \cite{chandler2005, leikin1993}, no interactions other than electromagnetic forces are available. An electromagnetic field is set up by dipole moments of water molecules and partial charges of molecules. In other words, we have a closed, smooth manifold in aqueous solution where charge and water molecules could migrate through the surface Fig.~\ref{fig:protein_DNA}. The surface can be of mixed nature (hydrophobic, hydrophilic or both) with randomly distributed polar or non-polar groups, can be compressible, continuously deformable and permeable against water and ionic charges. At the nanometer scale, for small masses, potential energy can be electromagnetic only. Therefore we have potential energy density constructed from the electromagnetic tensor plus the term related to variation of charges as it is defined in (\ref{27}). Even though modeling of potential energy as electromagnetic interaction energy is fairly clear, the dependence of these interactions on the object's geometry is not. The geometry dependence becomes visible only after the complete formulation of the equations of motion (\ref{53}).

\section{Results and Discussions}

\subsection{Poisson-Boltzmann Equation}

To demonstrate effectiveness of (\ref{53}) let's discuss free motion of two manifolds embedded in three dimensional Euclidean space for the stationary surface in electrostatic field. Therefore we have the following conditions ${\bm{V}=0}$ stationary surface in electrostatics field where ${\bm{a}=0, \bm{j}=0}$, ${\bm{A}_\cdot=(-\varphi/c,\bm{0})}$, $\bm{J}^\cdot=(cQ,\bm{0})$ and ${\bm{\partial}=(0,\partial_x,\partial_y,\partial_z)}$. Then from second equation of (\ref{53}) with the precondition (\ref{46}), we find
\begin{equation}
cQ-\frac{1}{\mu_0}\partial_\beta F^{\beta 0}=0 \label{56}
\end{equation}
Taking into account the definition of electromagnetic tensor and that we discuss electrostatic field, partial derivative of the electromagnetic tensor in (\ref{56}) is $(1/c)\partial_\beta E^\beta$ and therefore
\begin{equation}
\partial_\beta E^\beta=c^2\mu_0Q \label{57}
\end{equation} 
By the definition of the electric field $E^\beta=-\partial^\beta\varphi$ and ${c^2\mu_0=1/\epsilon_0}$ so that (\ref{57}) transforms in 
\begin{equation}
\partial_\alpha\partial^\alpha\varphi=\nabla^2\varphi=-\frac{Q}{\epsilon_0} \label{58}
\end{equation}
The equation (\ref{58}) is generally known as Poisson-Boltzmann Equation in vacuum and was proposed to describe the distribution of the electric potential in solution in the direction of the normal to a charged surface \cite{gouy1910, chapman1913, davis1990}.

\begin{figure}
\includegraphics{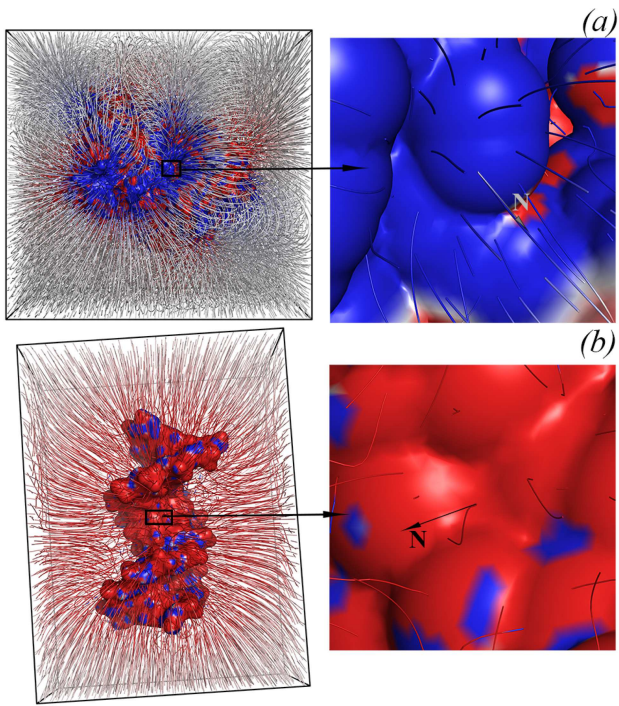}
\caption{\label{fig:electrostatics} (Color online) Color-coded electrostatic surface where red indicates negatively charged regions of the surface, white neutrally charged and blue positively charged one. Simulated electrostatic field lines are displayed as hairs on the surface of the protein $(a)$ and the DNA $(b)$. Hairs are generated by adaptive Poisson-Boltzmann solver and highlight how variations of the field lines on the surface describe the charge distribution pattern on the surface. Right side of the figure shows position of the surface normal $\bm{N}$ and electrostatic hairs.}
\end{figure}

Here we demonstrated that, the Poisson-Boltzmann Equation is a particular case and can be obtained from the equations of motion (\ref{53}) for stationary surfaces in electrostatic field. To support this statement we have generated electrostatic field lines by the Adaptive Poisson-Boltzmann Solver (APBS) \cite{dolinsky2007}  software for the protein \cite{svintradze2013structure} and the DNA \cite{drew1981} (Fig.~\ref{fig:electrostatics}).
As it is visible from the Fig.~\ref{fig:electrostatics} field lines follow the surface normal.

\subsection{Classical electrodynamics, Maxwell equations}
In this subsection we demonstrate that, the equation of motions simplify to Maxwell equations for stationary interfaces $C=0$ and massless $\rho=0$ three manifolds embedded in Minkowski space. Indeed, from the second and the third equations of (\ref{53}), taking into account that in stationary case second term in (\ref{32}) vanishes, we find  
\begin{align}
\int_\Omega\mathcal{F}\mathcal{A}d\Omega&=0 \label{59} \\
\int_\Omega\mathcal{F}^i\mathcal{A}_id\Omega&=0 \label{60}
\end{align}
Adding (\ref{59}) to (\ref{60}) and taking into account (\ref{47}) and (\ref{49}) one obtains 
\begin{equation}
\mathcal{F}\mathcal{A}+\mathcal{F}^i\mathcal{A}_i=\pmb{\mathscr{F}}\cdot\frac{\partial\bm{A}}{\partial t}=0 \label{61}
\end{equation}
(\ref{61}) must hold for any partial time derivative of the four vector potential, therefore
\begin{equation}
\pmb{\mathscr{F}}=(\mathcal{F}^\alpha)=(J^\alpha-\frac{1}{\mu_0}\partial_\beta F^{\beta\alpha})=0
\end{equation}
and the Maxwell equations with the source in the vacuum follows.
\begin{equation}
\frac{1}{\mu_0}\partial_\beta F^{\beta\alpha}=J^\alpha \nonumber
\end{equation}

We got somewhat unexpected result: any three manifold with stationary interface $C=0$ and with massless surface mass density $\rho=0$,\footnote{here surface mass density is same as the mass density of three manifold, because three manifold is the surface in 4D space.} satisfies Maxwell equations. However, arguably photon is only massless particle which satisfies Maxwell equation, therefor photon can be interpreted as stationary interface three manifold embedded in Minkowski space with vanishing surface mass density.

\subsection{Classical hydrodynamics, Euler equation}

In this section we simplify equations of motion using physical arguments and demonstrate that the equation system (\ref{53}) yields the Euler equation for dynamic fluid for some simplified cases. Indeed, let's propose that moving fluid has planar surface $B_{ij}=0$ with stationary interface $C=0$ and is embedded in Euclidean three space, then simplifications of (\ref{53}) lead to a system of equations of motion
\begin{align}
\dot{\nabla}\rho+\nabla_i(\rho V^i)&=0 \nonumber \\
\mathcal{F}\mathcal{A}&=0  \nonumber \\
\int_S\rho V_i(\dot{\nabla}V^i+V^j\nabla_jV^i)dS &=\int_\Omega\mathcal{F}^i\mathcal{A}_id\Omega \label{63}
\end{align}
The first equation of (\ref{63}) is continuity equation for the surface mass density and is conservation of mass at the flat space, the second one yields that normal component of the dot product ${\pmb{\mathscr{F}}\cdot(\partial\bm{A}/\partial t)}$ vanishes. To simplify the last equation of (\ref{63}) we note that the total 'force' acting on the volume is equal to the integral $-\int_SpdS$ of the total pressure $p$, taken over the boundary (surface) of the volume. Applying the Gauss theorem to the surface integral by taking into account that pressure across the surface acts on normal direction so that it can be written as $p=p_\alpha N^\alpha$, then
\begin{equation}
-\int_SpdS=-\int_Sp_\alpha N^\alpha dS=-\int_\Omega\nabla^\alpha p_\alpha d\Omega \label{64}
\end{equation} 
On the other hand $\mathcal{F}^i\mathcal{A}_i$ is a cause of the gradient of the tangential velocity and the tangential gradient of the pressure, therefor
\begin{equation}
\mathcal{F}^i\mathcal{A}_i=-\nabla^\alpha(V_i\nabla^i p_\alpha) \label{65}
\end{equation}
Taking into consideration (\ref{64}, \ref{65}) in (\ref{63}) and applying Gauss theorem to the space integral, we find
\begin{align}
\int_S\rho V_i(\dot{\nabla}V^i+V^j\nabla_jV^i)dS &=\int_\Omega\mathcal{F}^i\mathcal{A}_id\Omega \nonumber \\
&=-\int_\Omega\nabla^\alpha(V_i\nabla^i p_\alpha)d\Omega \nonumber \\
&=-\int_SV_iN^\alpha\nabla^i p_\alpha dS \label{66}
\end{align}
According to Weingarten's formula (\ref{24}) $N^\alpha$ is invariant vs the surface derivative for flat manifolds and, therefore, can be taken into the surface covariant derivative, so that $V_iN^\alpha\nabla^i p_\alpha=V_i\nabla^i p$.\footnote{more information about how the term $\pmb{\mathscr{F}}\cdot\partial\bm{A}/\partial t$  can be modeled as gradient of pressure times velocity can be found in \cite{svintradze2016}.} Then (\ref{66}) after subtracting $V_i$ yields
\begin{equation}
\rho(\dot{\nabla}V^i+V^j\nabla_jV^i)=-\nabla^i p \label{67}
\end{equation}
Taking into account that for flat surfaces ${\dot{\nabla}=\partial/\partial t}$ and ${\nabla_j=\partial_j}$ one immediately recognizes the last equation (\ref{67}) is the exact analog of the classical Euler equation of fluid dynamics.

As we stated above the equations of motion (\ref{53}) is formulated for freely moving manifolds i.e. interaction with environment is ignored and the matter is set to be a vacuum. Though it can be trivially generalized for the matter and then simplifications, instead of giving Euler equation, will lead to more complete Navier-Stokes equation and or magnetohydrodynamic equations. For instance, in the matter, according to (\ref{55}), electromagnetic tensor becomes sum of electric displacement and magnetization tensors. Therefor, in (\ref{67}), instead of pure pressure gradient we will have additive term coming from magnetic field so that (\ref{67}) will transform in ideal magneto hydrodynamic equation. 

Analogically, if interaction with an environment is taken into account, then instead of single surface we have two surfaces at the surface/environment interface and the Lagrangian (\ref{30}) is split by two kinetic energy terms, one for surface and another one for environmental interface. All these will rise additive terms in the third equation of (\ref{53}) so that the equation (\ref{67}) will transform in Navier-Stokes equation.

\subsection{Equilibrium shapes of micelles}

Let's answer the question: what is a shape of micelles, formed from lipid molecules, when they are in mechanical equilibrium with solvent. Lipids have hydrophilic heads and hydrophobic tails, so that, in solutions, they tend to form the surface with heads on one side and tails on the other. Since the tails disperse the water molecules, the surface made is closed and has some given volume. Such structures are called micelles \cite{tanford1973}. Since lipids form a homogeneous surface, in equilibrium conditions we must have 
\begin{equation}
\mathcal{F}\mathcal{A}=const\neq 0 \label{68}
\end{equation}
and $\mathcal{F}^i\mathcal{A}_i=0$. Usually speed of micelle interface dynamic is in the range of $~nm/ns$ and, therefore, there is no necessity of discussion relativistic formalism so that the surface is two dimensional and the space is Euclidean. As far as the surface dynamics is slow magnetic field is much smaller then electric field $B^2<<E^2$ and the potential energy becomes
\begin{equation}
 \mathcal{U}=-\int_\Omega (\frac{\epsilon_0}{2}E^2+\varphi Q)d\Omega \label{69}
\end{equation}  
Using first low of thermodynamics, (\ref{69}) can be modeled as volume integral from the surface pressure \cite{svintradze2016}, therefore 
\begin{equation}
{p=\frac{\epsilon_0}{2}E^2+\varphi Q} \label{70}
\end{equation}
On the other hand, taking into account the conditions (\ref{68},\ref{69}), the total potential energy of the surface can be modeled as 
\begin{equation}
\mathcal{U}=\sigma\int_SdS \label{71}
\end{equation}
Taking into account (\ref{71}), the system Lagrangian becomes same as it is in (\ref{1}) and its variation leads to the equation
\begin{equation}
\rho (\dot{\nabla}C+2V^i\nabla_iC+V^iV^jB_{ij})=\sigma B_i^i \label{72}
\end{equation}
(\ref{72}) was first reported in \cite{grinfeld2009}. Using (\ref{69},\ref{70},\ref{72}) in the equations of motions (\ref{53}), after simple algebra we fined
\begin{equation}
\partial_\alpha (\sigma V^\alpha B_i^i+pV^\alpha)=-V^\alpha\partial_\alpha p \label{73}
\end{equation}
When the homogeneous surface, such is micelle, is in equilibrium with the environment then the solution of the (\ref{73})\footnote{shorter alternative way to deduce (\ref{73}) and about it's solution in equlibrium conditions is given in \cite{svintradze2016}.} is
\begin{equation}
B^i_i=-\frac{p}{\sigma} \label{74}
\end{equation}
From the equation (\ref{74}) immediately follows generalized  Young-Laplace relation which connects surface pressure to Gaussian curvature and surface tension. (\ref{74}) dictates that the homogeneous surfaces in equilibrium with environment adopts the shape with constant mean curvatures (CMC), therefore explains well anticipated lamellar, cylindrical and spherical shapes of micelles. This is another unexpected and surprisingly simple solution to the equations of motions (\ref{53}).

\section{conclusions}

We have proposed equations of moving surfaces in an electromagnetic field and demonstrated that the equations simplify to: 1) Maxwell equations for massless three manifolds with stationary interfaces; 2) Euler equations for dynamic fluid for planar two manifolds with stationary interface embedded in Euclidean space which can be generalized to Navier-Stokes equations and to magneto-hydrodynamic equations; 3) Poisson-Boltzmann equation for stationary surfaces in electrostatic field.

We have applied the equation to analyze motion of hydrophobic-hydrophilic surfaces and explained 'equilibrium' shapes of micelles. The application was done on a protein, DNA dodecamer and micelles. In all cases analyses were in good qualitative as well as quantitative agreement with known experimental results for micelles \cite{svintradze2016} and with simulations for the protein and the DNA. Analytic solutions to simplified equations for homogeneous surfaces in equilibrium with environment produced generalized Young-Laplace law and explained why mean curvature surfaces are such abundant shapes in nature.

Also we have showed that hydrophobic-hydrophilic effects are just another expression of well known electromagnetic interactions. In particular, equations of motion for moving surfaces in hydrophobic and hydrophilic interactions, together with the  analytic solution, provide an explanation for the nature of the hydrophobic-hydrophilic effect. Hydrophobic and hydrophilic interactions are dispersive interactions throughout the molecules and conform to electromagnetic interaction's dependence on surface morphology of the material bodies.

\begin{acknowledgments}
We were partially supported by personal savings accumulated during the visits to Department of Mechanical Engineering, Department of Chemical Engineering, OCMB Philips Institute and Institute for Structural Biology and Drug Discovery of Virginia Commonwealth University in 2007-2012 years. We thank Dr. Alexander Y. Grosberg from New York University for comments on an early draft of the paper and Dr. H. Tonie Wright from Virginia Commonwealth University for editing the English of an early draft of the paper. Limited access to Virginia Commonwealth University's library in 2012-2013 years is also gratefully acknowledged.

\end{acknowledgments}

\bibliographystyle{apsrev4-1} 
\bibliography{References}% Produces the bibliography via BibTeX.

\end{document}